\documentclass[12pt]{article}
\pdfoutput=1
\usepackage[totalwidth=480pt,totalheight=640pt]{geometry}
\usepackage[centertags]{amsmath}
\usepackage[square, comma, sort&compress,numbers]{natbib}
\usepackage{array,multirow}
\usepackage{amssymb}
\usepackage{graphicx}
\usepackage{color}
\usepackage{setspace}

\usepackage{epsfig}

\def\Or[#1]{{\text{O}}\left({#1}\right)}
\def\dotl[#1,#2]{\left\langle #1, #2 \right\rangle}
\def\dotlb[#1,#2]{[ #1, #2 ]}
\def\dotp[#1,#2]{(#1) \cdot (#2)}
\def\aff[#1,#2]{\hat{#1}(#2)}
\def\n4sym{{\cal N}=4 SYM}
\def\>{\rangle}
\def\<{\langle}
\def\weight[#1,#2,#3]{\{(#1),#2,#3\}}
\def\ads[#1]{$\text{AdS}_{#1}$}

\newcommand{\ba}{\begin{eqnarray}}
\newcommand{\ea}{\end{eqnarray}}
\newcommand{\be}{\begin{eqnarray}}
\newcommand{\ee}{\end{eqnarray}}

\newcommand{\CD}{{\cal D}}

\newcommand{\CG}{{\cal G}}

\newcommand{\CO}{{\cal O}}

\newcommand\oo\infty
\newcommand\s\sigma
\newcommand\de\delta
\newcommand\De\Delta
\newcommand\p[1]{\left(#1\right)}
\newcommand\f\phi
\newcommand\g\gamma
\newcommand\x\times
\usepackage{hyperref}

\begin{document}

\begin{titlepage}

\begin{center}
\vspace{1cm}

{\Large \bf  Conformal Blocks  in the Large $D$ Limit }

\vspace{0.8cm}

\small
\bf{A. Liam Fitzpatrick$^1$,  Jared Kaplan$^{1,2}$, David Poland$^3$}
\normalsize

\vspace{.5cm}

{\it $^1$ Stanford Institute for Theoretical Physics, Stanford University, Stanford, CA 94305}\\
{\it $^2$ Department of Physics and Astronomy, Johns Hopkins University, Baltimore, MD 21218} \\
{\it $^3$ Department of Physics, Yale University, New Haven, CT 06520} \\

\end{center}

\vspace{1cm}

\begin{abstract}

We derive conformal blocks in an inverse spacetime dimension expansion.  In this large $D$ limit, the blocks are naturally written in terms of a new combination of conformal cross-ratios.  We comment on the implications for the conformal bootstrap at large $D$.

\end{abstract}

\bigskip

\end{titlepage}

\section{Introduction and Review}
\label{sec:Intro}

Conformal Field Theories (CFTs) are important for many reasons: well-defined Poincar\'e invariant Quantum Field Theories can be viewed as renormalization group (RG) flows between conformal fixed points, numerous condensed matter systems are described by CFTs, and via the AdS/CFT correspondence \cite{Maldacena:1997re, Witten, Gubser:1998bc}, quantum gravity can also be approached as the study of CFTs.  Much is known about many classes of CFTs, especially weakly coupled CFTs, supersymmetric CFTs, 2D CFTs, and large $N$ CFTs. 
However, relatively little is known about the full space of unitary CFTs \cite{Douglas:2010ic}.    In particular, although the superconformal algebra exists only in $D \leq 6$ spacetime dimensions \cite{Kac:1977em, Nahm:1977tg} and we have examples of CFTs up to $D=6$, we do not know if non-trivial\footnote{Free theories are an example of trivial CFTs.  We will also view `mean field theories' or `generalized free theories', namely CFTs whose correlators are directly determined by two-point functions, as trivial CFTs, although these theories do not have an energy-momentum tensor. }  unitary CFTs exist in $D > 6$ dimensions. 

The purpose of this paper is to begin to study CFT in larger $D$ by analyzing the conformal blocks and the conformal bootstrap \cite{Polyakov:1974gs, Belavin:1984vu}, using $1/D$ as an expansion parameter.\footnote{Some other examples of a $1/D$ expansion include \cite{Strominger:1981jg, BjerrumBohr:2003zd, Hamber:2005vc, Emparan:2013moa, Emparan:2013xia,Prester:2013gxa}.}  As reviewed in section \ref{sec:BootstrapReview}, the conformal bootstrap depends only on the principles of conformal symmetry and unitarity, so it provides very general constraints on CFTs. Indeed, it has been demonstrated in a series of recent numerical studies~\cite{Rattazzi:2008pe, Rychkov:2009ij,Caracciolo:2009bx,Poland:2010wg,Rattazzi:2010gj,Rattazzi:2010yc,Vichi:2011ux,Poland:2011ey,ElShowk:2012ht,Liendo:2012hy,ElShowk:2012hu,Beem:2013qxa} that the bootstrap places extremely interesting constraints on CFTs in $D>2$.  One might hope that if non-trivial CFTs are prohibited in certain circumstances, such as at large $D$, then this may also be proven using the bootstrap.

The conformal blocks \cite{Polyakov:1974gs, Ferrara:1973vz, Ferrara:1973eg, Ferrara:1973yt, Ferrara:1974nf, Dolan:2000ut, Dolan:2003hv, Dolan:2011dv, Costa:2011dw, DSDProjectors} are the fundamental ingredients needed for the bootstrap.  They are atomic contributions to CFT four-point functions, corresponding to the exchange of a single irreducible representation of the conformal group between two pairs of CFT operators.  In general they are fairly complicated functions that can be understood and computed as eigenfunctions of the quadratic Casimir operator of the conformal group \cite{Dolan:2003hv}.  In this paper we will present a new pair of variables such that the conformal Casimir differential operator becomes separable at large $D$.  At leading order in $1/D$, the conformal blocks can then be written in closed form in terms of these variables.  

Using these leading order large $D$ conformal blocks and recent analytic methods from \cite{JP, Fitzpatrick:2012yx, Komargodski:2012ek}, one might hope to derive powerful analytic constraints on large $D$ CFTs.  
We will give some initial arguments along these lines, where one can see that anomalous dimensions must vanish in the strict $D = \infty$ limit.  On the other hand, the best finite $D$ constraints from the bootstrap that we have found so far already follow from the large spin results of \cite{Fitzpatrick:2012yx, Komargodski:2012ek}, combined with an expansion of the conformal blocks from \cite{Dolan:2011dv}. These results confirm, but do not extend beyond, expectations from effective field theory at large $D$, although it is interesting that they can be derived for general CFTs.

We will  be interested only in unitary CFTs, and this puts $D$-dependent restrictions on operator dimensions.  Specifically, in order to have positive-norm states we  require 
\be
\label{eq:UnitarityBound}
\Delta &\geq& \frac{D}{2} - 1  \ \ \ \mathrm{for} \ \ \ \ell = 0 , \\
\Delta &\geq& D - 2 + \ell \ \ \ \mathrm{for} \ \ \ \ell > 0 , 
\ee
so we must take all operator dimensions $\Delta \propto D$ in the large $D$ limit.  Thus, when we expand in $1/D$ we cannot neglect terms of order $\Delta / D$.  When we compute the conformal blocks we will also keep terms of order $\ell / D$, capturing  both the regime $\ell \ll D$ and $\ell \propto D$.  

As a guide for what to expect,  let us consider the behavior of effective field theories (EFTs) in large $D$, both in the CFT spacetime and also in a dual AdS description.  In EFT in $D$ dimensions any massless bosonic field $\phi$ should have an action of the schematic form
\be
S_{CFT} = \int d^D x \left[ (\partial \phi)^2 + \frac{1}{\Lambda^{\frac{D}{2} - 3} } \phi^3 + \cdots \right] ,
\ee
where the elipsis indicates higher order terms.  We immediately see that perturbative EFTs at large $D$ must flow to a free fixed point in the IR.  At low-energies near this fixed point, the interactions will be exponentially small in $D$, since they will be roughly proportional to a power of $(E/\Lambda)^{D}$.  We can attempt to find more interesting CFTs by studying EFT in AdS space for a bulk field $\Phi$ coupled to gravity, with an action
\be
S_{AdS} = \int d^{D+1} X \sqrt{-g} \left[ M_{pl}^{D-2} R + (\partial \Phi)^2 + \frac{1}{\Lambda^{\frac{D-5}{2}  } } \Phi^3 + \cdots \right] .
\ee
By using the AdS/CFT dictionary we can use this action to obtain approximate CFT correlators that describe the low scaling-dimension spectrum of the CFT \cite{Katz, ElShowk:2011ag, Fitzpatrick:2012cg}.  This greatly increases the space of possibilities that we can consider for the CFT correlators.  However, in this case the CFT will again be nearly Gaussian, in the sense that all its correlators are determined by two-point functions up to corrections that are exponentially small in $D$.  In particular, this means that all anomalous dimensions and corrections to OPE coefficients for the low-dimension operators in the CFT will be exponentially small in $D$ in the regime where the EFT description makes sense.  
Not all CFTs are known to have local Lagrangian descriptions, so while the previous arguments may provide strong expectations for the structure of CFTs at large $D$, they are necessarily limited in their applicability. 
Analysis based on the conformal bootstrap is more robust.  We will be able to argue using the bootstrap that anomalous dimensions at large spin are necessarily exponentially suppressed, a result following from the analysis of \cite{Fitzpatrick:2012yx, Komargodski:2012ek}.

The outline of this note is as follows.  In the next section we will briefly review the conformal bootstrap.  Then in section \ref{sec:ConformalBlocks} we will derive expressions for the conformal blocks in the large $D$ limit.  We begin with scalar conformal blocks in order to motivate a new set of variables, then we use the conformal Casimir differential equation to compute more general blocks.  In section \ref{sec:ObservationsImplications} we compare these blocks to their finite $D$ counterparts and make some relevant observations about their analytic structure.  Finally, we discuss some implications in the context of the bootstrap and mention some possibilities for future work.

\subsection{Brief Bootstrap Review}
\label{sec:BootstrapReview}

In CFTs, the bootstrap condition follows from the constraints of conformal invariance combined with the operator product expansion (OPE), which says that a product of local operators is equivalent to a sum
\be
\phi(x) \phi(0) &=& \sum_{\CO} \lambda_{\CO} C^I_{\CO}(x,\partial) \CO_I(0).
\ee
  Conformal symmetry relates the coefficients of all operators in the same irreducible conformal multiplet, and this allows one to reduce the sum above to a sum over different irreducible multiplets. Each of these multiplets is labeled by a lowest-weight (i.e., ``primary") operator, which is annihilated by the special conformal generator.  When this expansion is performed inside of a four-point function, the contribution of each block is just a constant ``conformal block coefficient'' $P_{\CO} \propto \lambda_{\CO}^2$ for the entire multiplet times a function of the $x_i$'s whose functional form depends only on the spin $\ell_{\CO}$ and dimension $\Delta_{\CO}$ of the primary operator:
  \be
  \< \phi(x_1) \phi(x_2) \phi(x_3) \phi(x_4) \> &=&  \frac{1}{(x_{12}^2 x_{34}^2)^{\Delta_{\phi}}} \sum_{\CO} P_{\CO} G^{(D)}_{\Delta_\CO, \ell_{\CO}}(u,v),
  \ee
where $x_{ij} \equiv x_i - x_j$,  
and 
\be
u= \left( \frac{x_{12}^2x_{34}^2}{ x_{13}^2 x_{24}^2 } \right), \qquad  v = \left( \frac{x_{14}^2 x_{23}^2}{x_{13}^2 x_{24}^2 } \right),
\ee
 are the conformally invariant cross-ratios.  The functions $G^{(D)}_{\Delta_\CO, \ell_{\CO}} (u,v)$ are usually referred to as conformal blocks or conformal partial waves \cite{Polyakov:1974gs, Ferrara:1973vz, Ferrara:1973eg, Ferrara:1973yt, Ferrara:1974nf,Dolan:2000ut, Dolan:2003hv, Dolan:2011dv, Costa:2011dw, DSDProjectors}, and they are crucial elementary ingredients in the bootstrap program.
 
 In the above, we took the OPE of $\phi(x_1) \phi(x_2)$ and $\phi(x_3) \phi(x_4)$ inside the four-point function, but one can also take the OPE in the additional ``channels'' $\phi(x_1) \phi(x_3)$ and $\phi(x_2) \phi(x_4)$ or $\phi(x_1) \phi(x_4)$ and $\phi(x_2) \phi(x_3)$. The bootstrap equation is then the condition that the decompositions in different channels agree:
 \be
 \label{eq:bootstrap}
   \frac{1}{(x_{12}^2 x_{34}^2)^{\Delta_{\phi}}} \sum_{\CO} P_{\CO}G^{(D)}_{\Delta_\CO, \ell_{\CO}} (u,v) &=&
     \frac{1}{(x_{14}^2 x_{23}^2)^{\Delta_{\phi}}} \sum_{\CO} P_{\CO}G^{(D)}_{\Delta_\CO, \ell_{\CO}} (v,u).
\ee
Much of the power of this constraint follows from the fact that by unitarity, the conformal block coefficients $P_{ \CO}$ must all be non-negative in each of these channels, because the $P_{\CO}$ can be taken to be the squares of real OPE coefficients.

\section{Conformal Blocks at Large Spacetime Dimension}
\label{sec:ConformalBlocks}

In this section we will compute expressions for the conformal blocks in the limit that the spacetime dimension $D$ is large.  First we will compute the scalar blocks as a suggestive warm-up; this can be done by taking the large $D$ limit of an expansion of the scalar block in a double power series which holds for all $D$.  Then we will generalize these blocks to arbitrary intermediate spin by studying the large $D$ limit of the conformal Casimir differential equation.  For this purpose we will transform to a new set of variables
 \be
 y_{\pm} \equiv  \frac{u}{(1 \pm \sqrt{v})^2},
 \ee
suggested by our first result for the scalar blocks.  
 We find that in terms of these variables, the Casimir differential equation separates at large $D$ and can be explicitly solved in terms of ${}_2 F_1$ hypergeometric functions.  In the last subsection we make some observations about these results and provide some tests of their validity.

\subsection{Scalar Block Warm-Up}

To begin, we will make use of the double power series expansion for the scalar conformal blocks in any number of spacetime dimensions \cite{Dolan:2000ut, Dolan:2011dv}
\be\label{eq:series}
G_{\Delta,0}^{(D)}(u,v) &=& \sum_{m,n=0}^{\infty} \frac{(\frac{\Delta}{2})_n^2 (\frac{\Delta}{2})_{n+m}^2}{(\Delta+1-\frac{D}{2})_n (\Delta)_{2n+m}} \frac{u^{\frac{\Delta}{2}+n}}{n!} \frac{(1-v)^m}{m!} ,
\ee
where $u$ and $v$ are the conformal cross-ratios.  This can be summed exactly in $m$ to give
\be\label{eq:msum}
G_{\Delta,0}^{(D)}(u,v) &=& \sum_{n=0}^{\infty} \frac{(\frac{\Delta}{2})_n^4}{(\Delta+1-\frac{D}{2})_n (\Delta)_{2n}} {}_2F_1\left(\frac{\Delta}{2}+n,\frac{\Delta}{2}+n,\Delta+2n; 1-v \right) \frac{u^{\frac{\Delta}{2}+n}}{n!} .
\ee
Now let us try to understand the large $D$, $\Delta$ limit of these scalar blocks.  First, note that a hypergeometric function ${}_2F_1(a,a;2a;x)$ can be approximated at large $a$ as:
\be
 \left(\frac{\Gamma(a)^2}{\Gamma(2a)}\right) {}_2F_1(a,a;2a;x) &=& \int_{0}^1 \frac{dt}{t(1-t)} \left( \frac{t(1-t)}{1-t x} \right)^a 
\nonumber \\ &\approx& \sqrt{\frac{\pi}{a}}\frac{1}{(1-x)^{1/4} (1+\sqrt{1-x})^{2a-1}}.
\ee
This saddle point approximation was discussed more extensively in Appendix A of~\cite{Fitzpatrick:2012yx}.

Inserting this approximation, we would like to do the sum 
\be
G_{\Delta,0}^{(D)}(u,v) &\approx&  \sqrt{\pi} \frac{1+\sqrt{v}}{v^{1/4}} \frac{\Gamma(\Delta)}{\Gamma(\frac{\Delta}{2})^2} \sum_{n=0}^{\infty} \frac{(\frac{\Delta}{2})_n^2}{(\Delta+1-\frac{D}{2})_n} \frac{1}{\sqrt{n+\frac{\Delta}{2}}} \frac{\left(\frac{u}{(1+\sqrt{v})^2}\right)^{\frac{\Delta}{2}+n}}{ n!} ,
\ee
in the large $D$, $\Delta$ limit.  In this limit one can approximate
\be
\frac{1}{\sqrt{z}} \approx \frac{\Gamma(z+a)}{\Gamma(z+a+\frac{1}{2})}
\ee
for any $a \ll z$, where $z = n + \frac{\Delta}{2} \gg 1$.   Taking $a=-1/2$, this gives
\be
G_{\Delta,0}^{(D)}(u,v) &\approx&  \sqrt{\pi} \frac{1+\sqrt{v}}{v^{1/4}} \frac{\Gamma(\Delta)}{\Gamma(\frac{\Delta}{2})^2} \sum_{n=0}^{\infty} \frac{(\frac{\Delta}{2})_n^2}{(\Delta+1-\frac{D}{2})_n} \frac{\Gamma(n + \frac{\Delta}{2} - 1/2)}{\Gamma(n + \frac{\Delta}{2})} \frac{y_+^{\frac{\Delta}{2}+n}}{ n!} ,
\nonumber \\ &\approx& 
2^{\Delta-1} \frac{1+\sqrt{v}}{v^{1/4}}  y_+^{\frac{\Delta}{2}} {}_2F_1\left(\frac{\Delta-1}{2},\frac{\Delta}{2};\Delta-\frac{D}{2}+1; y_+ \right) ,
\ee
where $y_+ = \frac{u}{(1+\sqrt{v})^2}$.  This approximation should be very good in the limit where $\Delta \propto D \to \infty$, as is required by unitarity.

One can go on to study blocks with spin $\ell > 0$ by using recursion relations that relate blocks of different spins \cite{Dolan:2011dv}.  This leads to the result that for $\ell \ll D$, higher spin blocks are related to lower spin blocks by multiplication with $1/y_-$, where $y_- = \frac{u}{(1-\sqrt{v})^2}$.  Instead of performing this analysis, let us move on to study the conformal Casimiar differential equation, which will allow for a general solution for any $\Delta$ and $\ell$.

\subsection{ Solving the Conformal Casimir Differential Equation }

Conformal blocks are solutions of the conformal Casimir differential equation in two variables \cite{Dolan:2003hv}.  This equation is usually written in terms of the conformal cross-ratios $u$ and $v$ or the variables $z$ and $\bar z$, defined by $z \bar z = u$ and $(1-z)(1- \bar z) = v$.  We will write the Casimir differential equation in terms of the new set of variables $y_+ \equiv  \frac{u}{(1+\sqrt{v})^2}$ and $y_- \equiv \frac{u}{(1-\sqrt{v})^2}$, which have some similarity to the variables recently employed by \cite{Hogervorst:2013sma}.  We saw above that $y_+$ arises naturally in the case of the large $D$ scalar conformal block.  

The differential equation takes the form
\be
\mathcal{D} G_{\Delta,\ell}^{(D)} = \frac12 C_{\Delta,\ell}^{D} G_{\Delta,\ell}^{(D)} ,
\ee
where the conformal Casimir eigenvalue is given by $C_{\Delta,\ell}^{D} = \Delta(\Delta-D) + \ell(\ell+D-2)$.  In terms of $y_\pm$, the differential operator $\mathcal{D}$ is:
\be
\mathcal{D} = \mathcal{D}_{y_+} + \mathcal{D}_{y_-} + \frac{2}{y_+-y_-} \left(y_+^2(1-y_+)\partial_{y_+} - y_-^2(1-y_-) \partial_{y_-}\right)
\ee
with
\be
\CD_y \equiv 2y^2(1-y)\partial_y^2 - y(y+D-2)\partial_y .
\ee
The $\CD_y$ operators are generically  of order $D^2$ when acting on conformal blocks.  The advantage of these variables is that the mixed term in the full differential operator $\CD$ is only of order $D$, and therefore is subleading in the large $D$ limit.  This approximation breaks down when $y_+ - y_- \lesssim \frac{1}{D}$; this is the region where the conformal cross-ratio $v \lesssim \frac{1}{D^2}$.  The region where $v \ll 1$ played a crucial role in the analysis of \cite{Fitzpatrick:2012yx}, and it will also be important for us later on when we discuss the bootstrap.

Writing $G_{\Delta,\ell}^{(D)} = \frac{1}{\sqrt{y_- - y_+}} H_{\Delta,\ell}^{(D)}$, we find the simpler differential equation
\be
\left[\mathcal{D}_{y_+} + \mathcal{D}_{y_-} - \frac{y_+ y_-}{2(y_+ - y_-)^2} (y_+ + y_- - 2) \right] H_{\Delta,\ell}^{(D)} = \frac12 (C_{\Delta,\ell}^{D} - D +1) H_{\Delta,\ell}^{(D)}.
\label{eq:largeDcasimirEq}
\ee
Let us assume that we can neglect the third (mixed) term on the left hand side in the large $D$ limit, which holds when $v \gg \frac{1}{D^2}$.  Then we can find a separable solution:
\be
H_{\Delta,\ell}^{(D)} = A_\Delta (y_+ ) A_{1- \ell} (y_-),
\ee
where
\be
\mathcal{D}_y A_\Delta(y) &=& \frac12 (\Delta)(\Delta-D) A_\Delta(y), \\
\mathcal{D}_{y} A_{1 - \ell} (y) &=& \frac12 (\ell-1)(\ell+D-1) A_{1-\ell} (y).
\ee

These have the solutions:
\be
A_\Delta(y) &=& y^{\Delta/2} {}_2F_{1}\left(\frac{\Delta-1}{2},\frac{\Delta}{2},\Delta-\frac{D}{2}+1;y\right), \\
A_{1-\ell}(y) &=& y^{(1-\ell)/2} {}_2F_{1}\left(-\frac{\ell}{2},\frac{1-\ell}{2},-\ell-\frac{D}{2}+2; y \right) .
\ee

There is also the issue of boundary conditions: the behavior of the conformal blocks is constrained by the OPE at $u \rightarrow 0 $ and $v\rightarrow 1$ with $u \ll (1-v)^2$ to be
\be
G_{\Delta,\ell}^{(D)} \approx (1-v)^{\ell} u^{\frac{\Delta-\ell}{2}} &\stackrel{y_+ \ll y_- \ll 1}{\approx} &2^{\Delta+\ell} y_+^{\Delta/2} y_-^{-\ell/2},
\ee
where we have used the fact that the limit  $y_+ \ll y_- \ll 1$ corresponds to $u \ll (1-v)^2 \ll 1$.  Our solution
can easily be seen to satisfy this condition. This also fixes our normalization convention for the blocks.
At small $\ell$, one can verify that the $\ell$ dependence is consistent with the recursion relations of \cite{Dolan:2011dv} relating different values of $\ell$.  Thus,
\be
\label{eq:FinalLargeSpinBlock}
\CG^{(D)}_{\Delta,\ell} &\equiv&  \frac{2^{\Delta + \ell }}{\sqrt{y_- - y_+}} A_\Delta(y_+) A_{1-\ell}(y_-) 
\ee
is the correct block in the limit of large $D$.

The OPE limit is in fact slightly more general than we considered above, and elucidates the appearance of the $A(y)$ functions.  More generally, the OPE limit requires only $z, \bar{z} \ll 1$  but with $z/\bar{z}$ arbitrary, or equivalently $u\sim 0$ and $v\sim 1$ but with $\frac{u}{(1-v)^2}$ arbitrary.  Writing $z=r e^{i \theta}, \bar{z} = r e^{-i \theta}$, it is natural to parameterize this free direction by the angular variable $\sigma^2 \equiv \cos^2 \theta \approx \frac{1}{y_-}$.  So, the OPE limit is enough to control the blocks at $y_+ \ll 1$ for any $y_-$.  In \cite{Dolan:2011dv}, it was shown that the dependence on $\sigma^2$ in the OPE limit is given, as one might expect, by a Gegenbauer polynomial $C^{(\frac{D-2}{2})}_\ell(\sigma)$:
\be
G_{\Delta,\ell}^{(D)} \ \sim \ u^{\frac{\Delta}{2}} 
 C^{(\frac{D-2}{2})}_\ell(\sigma) 
 \ \propto \ y_+^{\frac{\Delta}{2}} y_-^{-\frac{1}{2}} A_{1-\ell}(y_-),
\ee 
where we have taken $y_+ \ll 1$ and neglected constant coefficients. In other words, in the OPE limit, 
$y_+$ is a radial variable, $y_-$ is an angular variable, and the $A_{1-\ell}(y)$ function arises because the dependence on $y_-$ is determined for generic values and is just a Gegenbauer polynomial.

\section{Observations and Implications}
\label{sec:ObservationsImplications}

Let us begin by providing some tests of these blocks.  First we will compare them to known examples in low $D$.  Then we will show (numerically) that the blocks satisfy the bootstrap equation when we use mean field theory OPE coefficients, which are known in general $D$ \cite{unitarity}.    Finally, we will comment on the behavior of the blocks in the `eikonal' limit recently studied in \cite{Fitzpatrick:2012yx}, where $u \ll v \ll 1$.  We will see that once the limit $D \to \infty$ is taken, the conformal blocks no longer contain $\log v$ singularities at small $v$, shown in~\cite{JP, Fitzpatrick:2012yx, Komargodski:2012ek} to be related to perturbative anomalous dimensions at large spin.  However, as we will discuss, this is an artifact of taking the large $D$ limit before the small $u, v$ limit.  Finally, we discuss some other implications and possibilities for future work.

\begin{figure}[t!]
\begin{center}
\includegraphics[width=\textwidth]{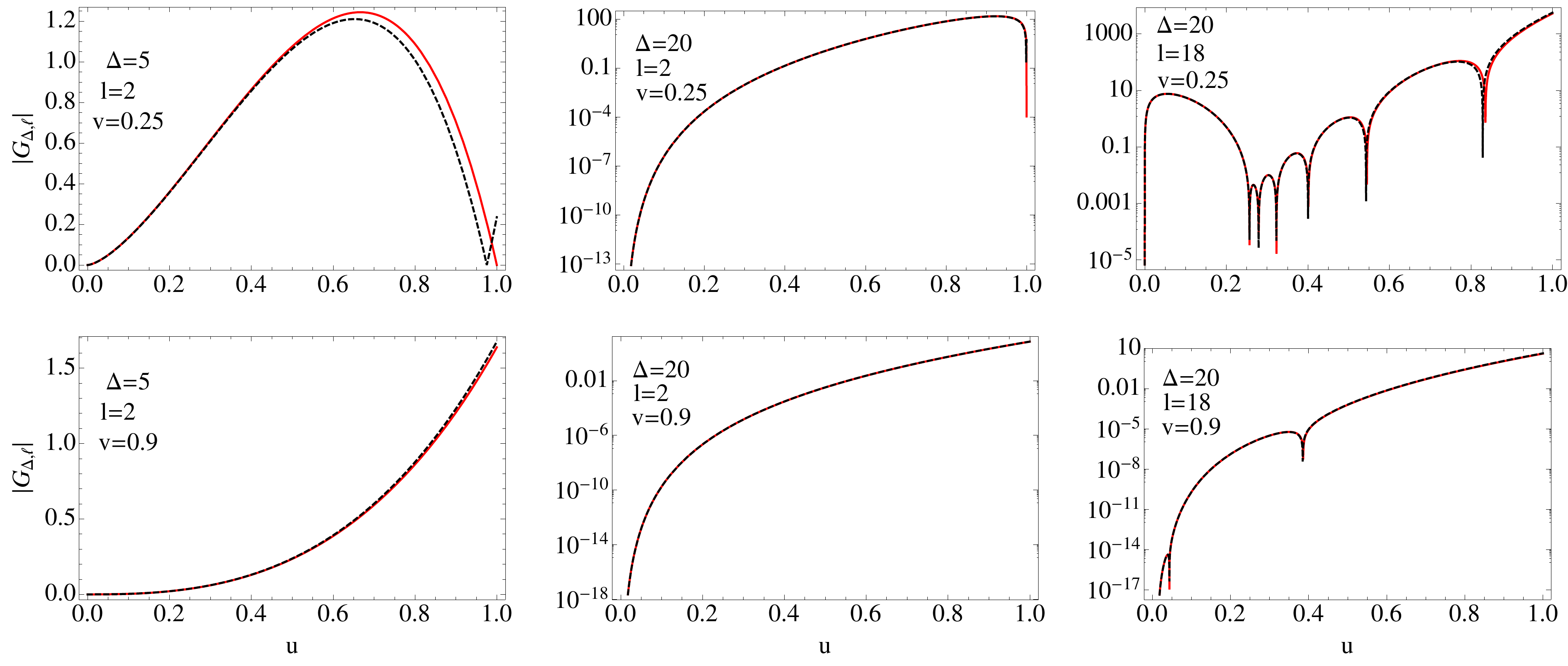}
\caption{Comparison of the absolute value of the exact conformal blocks $G^{(D)}_{\Delta,\ell}$ in $D=4$ (black, dashed line) with the large $D$ approximation formula $\CG^{(D)}_{\Delta,\ell}$ evaluated at $D=4$ (red, solid line).  Plots are shown as a function of the cross-ratio $u$ at fixed values of $v$.  One can generate similar plots and find what appears to be excellent agreement even when $D=2$.}
\label{fig:4dBlocksComparison}
\end{center}
\end{figure}

\subsection{Observations}

It is very straightforward to make a numerical comparison between our large $D$ blocks and known formulas at finite $D$.  Some direct comparisions are shown in figure \ref{fig:4dBlocksComparison}.  The large $D$ blocks appear to agree surprisingly well even for $D=4$, and in particular one can see that the non-trivial features are reproduced very precisely.  Generating similar plots comparing the large $D$ blocks to the exact blocks at $D=2$ shows similarly impressive and surprising agreement. This suggests that our simple large $D$ blocks might provide a useful approximation for further bootstrap studies.

As a more serious test, we can study the behavior of sums of blocks multiplied by conformal block coefficients.   In \cite{unitarity} the conformal block coefficients for mean field theory (or `generalized free theory', where all correlators follow from $\langle \phi(x) \phi(0) \rangle = x^{-2 \Delta_\phi}$) were determined for general $D$, they are 
\be
\label{eq:MFTCoeffs}
P_{n, \ell} = 
\frac{  \left[1+(-1)^{\ell} \right] (\Delta_\phi -h+1)_n^2 
   (\Delta_\phi)_{\ell+n}^2}{\ell! n! (\ell+h)_n (2 \Delta_\phi + n-2h+1)_n (2 \Delta_\phi + 2 n+\ell-1)_l (2 \Delta_\phi + n+\ell-h)_n} 
\ee
for external scalar operators with dimensions $\Delta_\phi$, where $h = D/2$.    Using these coefficients we can construct the crossing function
\be
F(u , v) \equiv \sum_{\Delta, \ell} {P_{\Delta, \ell} } \frac{v^{\Delta_\phi} G_{\Delta,\ell}^{(D)}(u,v) - u^{\Delta_\phi} G_{\Delta,\ell}^{(D)}(v,u) }{u^{\Delta_\phi} - v^{\Delta_\phi}} ,
\ee
where all $\Delta = 2 \Delta_\phi + 2n + \ell$ in the sum.  In this language, the conformal bootstrap reduces to the equation $F(u,v) = 1$.  Usually we use this equation to test if a certain choice of CFT spectrum and conformal block coefficients satisfies the constraint of crossing symmetry.  

\begin{figure}[t!]
\begin{center}
\includegraphics[width=0.84\textwidth]{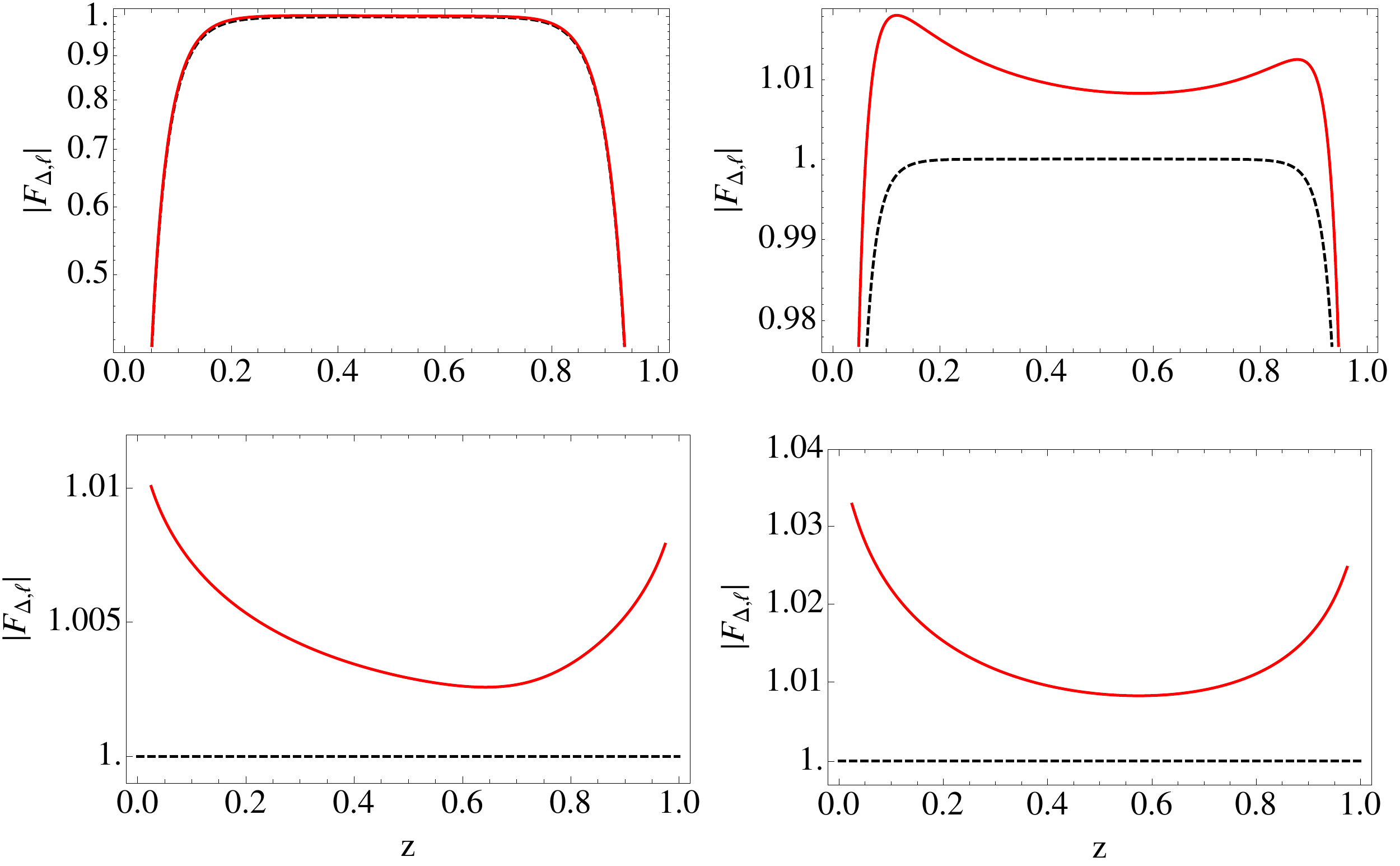}
\caption{These plots show partial sums of the absolute value of the bootstrap crossing function $F(z, \bar z)$ with the mean field theory conformal block coefficients from equation (\ref{eq:MFTCoeffs}),   comparing the exact 4D conformal blocks (black, dashed line) with the large $D$ conformal blocks evaluated at $D=4$ (red, solid line), as a function of $z$ with $\bar z = 0.3$.  In the plots on the left (right) we have $\Delta_e = 6.2 (2.2)$. 
The plots on the top have partials sums up to a maximum $n_{\rm max} = 10$ and $\ell_{\rm max} = 14$. The plots on the bottom have $n_{\rm max}$ and $\ell_{\rm max}$ large enough that the sums have numerically converged over the values plotted, so $F(z,\bar{z})=1 $ everywhere for the exact 4D conformal blocks.  In all cases we see extremely good agreement considering that here the large $D$ blocks neglect terms formally of $O(1/4)$.  }
\label{fig:4dSumComparison}
\end{center}
\end{figure}

However, in this context we can use our knowledge of the coefficients from equation (\ref{eq:MFTCoeffs}) to test the behavior of the large $D$ conformal blocks and compare them to the exact blocks.  A few examples in $D=4$ are shown numerically in figure \ref{fig:4dSumComparison}.  Although the agreement is not perfect, it again appears to be surprisingly good considering that the expansion is in $1/4$.  Note that agreement is not as good near $z = 0$ and $z=1$ so that $u$ or $v$ are near $0$.  This should be expected beause we ignored terms in the conformal Casimir differential equation that grow large in this limit, as can be seen in equation (\ref{eq:largeDcasimirEq}).  As we will discuss further below, this region is associated with conformal blocks of large spin, and can be studied analytically.

\subsection{Implications and Discussion}

Above, we have computed the asymptotic behavior of the conformal blocks in a large $D$ approximation.  We have seen that they provide a good numerical approximation to the exact conformal blocks even at $D = 4$, and that at large $D$ the partial sums of these blocks appear to converge in mean field theory.   We would like to go further and use the bootstrap to say something non-perturbative about the existence of CFTs at large $D$.  Here we will be able to recover the expectations from effective field theory in a new way, but unfortunately we have not yet found a way to make a more powerful statement.

In the strict $D \to \infty$ limit one might argue that operators with large angular momentum cannot have anomalous dimensions, invoking the results of \cite{Fitzpatrick:2012yx, Komargodski:2012ek} (see also \cite{Alday:2007mf}).  To be precise, consider the OPE of any two scalar operators 
\be
\CO_1(x) \CO_2(0) = \sum_{\CO} C_I(x, \partial) \CO^{I}_{\Delta, \ell}(0)
\ee
We will first study the case where we take the $D \to \infty$ limit first, before any other limits.   It was proved in \cite{Fitzpatrick:2012yx, Komargodski:2012ek} that at large $\ell$ there must exist operators $\CO_{\Delta, \ell}$  with dimension $\Delta = \Delta_1 + \Delta_2 + 2n + \ell + \gamma(n, \ell)$ for integers $n$ and $\ell$, where $\gamma(n, \ell)$ can be viewed as a small anomalous dimension at large $\ell$.  The same analysis with our $D \to \infty$ conformal blocks would imply that $\gamma(n, \ell) = 0$ at large $\ell$, precluding these anomalous dimensions at large spin.  In fact, a less na\"ive analysis at finite $D \gg 1$ implies that  $\gamma(n, \ell) \to 0$ at a rate of order $\ell^{-D}$ as $\ell \to \infty$.

These results follow from a study of the crossing relation organized as a sum over twists $\tau\equiv \Delta-\ell$ and spin $\ell$ instead of over dimension $\Delta$ and spin $\ell$
\begin{equation}
\label{eq:CrossingBootstrap}
1 +
\sum_{\tau,\ell} P_{\tau+\ell, \ell} \ \!  G_{\tau+\ell,\ell}^{(D)}(u,v) 
=  \p{\frac{u}{v}}^{\Delta_\phi}\p{1+\sum_{\tau,\ell} P_{\Delta+\ell,\ell} \ \!  G_{\tau+\ell,\ell}^{(D)}(v,u)},
\end{equation}
in the limit $u \to 0$, as discussed in detail in \cite{Fitzpatrick:2012yx}.  In this limit the left-hand side is simply $1$, while the right-hand side appears to vanish.  In fact the infinite sum over $\ell$ in the conformal blocks on the RHS results in a $u^{-\Delta_\phi}$ factor as $u \to 0$, so that both sides agree.  

Reorganizing the bootstrap equation as a sum over $\tau$ and $\ell$ instead of $\Delta$ and $\ell$ makes its structure at small $u$ clearer. The subleading terms from the conformal block(s) with smallest $\tau$ on the LHS of equation (\ref{eq:CrossingBootstrap}) behave as $u^{\tau}(a \log v + b)$ when we first expand at small $u$ and then subsequently expand the leading term at small $v$.  The $a \log v$ term can only be matched on the RHS of the crossing relation by non-zero anomalous dimensions $\gamma(n, \ell)$ in the large $\ell$ limit.   In fact, we find that  $a = 0$ if we use our result from equation (\ref{eq:FinalLargeSpinBlock}) and take $D \to \infty$ before we take the limit of small $u, v$.  This would seem to imply that anomalous dimensions $\gamma(n, \ell)$ vanish at large $\ell$ when $D \to \infty$.  However, one must be more careful because making this argument requires probing the region $u,v \lesssim 1/D^2$ where our blocks are no longer a good approximation.

More generally, one can use the general $D$ expansion as $u \to 0$~\cite{Dolan:2011dv}
\be
G_{\tau+\ell,\ell}^{(D)}(u,v) \sim u^{\tau/2}(1-v)^{\ell} {}_2F_1(\tau/2+\ell,\tau/2+\ell,\tau+2\ell,1-v)
\ee
to see that $a \neq 0$ if we first take the limit of small $u$ with $D$ finite before taking the limit of large $D$.  The various limits do not commute, and we must only take $D \to \infty$ after taking the other limits in order to get the correct result.  From \cite{Alday:2007mf, Fitzpatrick:2012yx, Komargodski:2012ek} we have
\be
\gamma(n, \ell) = \frac{\gamma_n}{\ell^{\tau}} + \cdots
\ee
at large $\ell$, where $\tau$ is the minimum twist of the conformal blocks appearing on the LHS of equation (\ref{eq:CrossingBootstrap}).  These leading twist operators are generically conserved currents $J_\mu$ or $T_{\mu \nu}$ with $\tau = D - 2$, or else they are scalars with $\tau > D/2 - 1$ due to the unitarity bound of equation (\ref{eq:UnitarityBound}).  So we see that the anomalous dimensions cannot fall off faster than $\sim \ell^{-D}$ at large spin.    The anomalous dimensions are forced to vanish identically only if we take the wrong order of limits and set $D = \infty$ from the beginning.  

This analysis has reproduced expectations from effective field theory in AdS as discussed in the introduction: we have re-discovered the fact that the anomalous dimensions are exponentially small as a function of $D$ at large distances in AdS.  A more nuanced analysis will be required to arrive at a definitive conclusion about the existence of CFTs in general dimensions.  In the future it will be interesting to explore the large $D$ bootstrap further, perhaps using a numerical approach.  It will also be interesting to further investigate how the conformal blocks change with $D$, as there must be a great deal of physics hidden in the behavior of these functions.  Perhaps the large $D$ expansion, which seems to be a remarkably accurate approximation even for $D = 4$, may shed light on these issues even for conformal field theories in a small number of spacetime dimensions.

\section*{Acknowledgments}    

We are grateful to Juan Maldacena, Jo\~ ao Penedones, and Steve Shenker for discussions, and David Simmons-Duffin for many discussions and collaboration.  We would also like to thank the participants of the ``Back to the Bootstrap II" workshop for discussions and the Perimeter Institute for hospitality during the early stages of this work.   This material is based upon work supported in part by the National Science Foundation Grant No. 1066293. ALF and JK were partially supported by ERC grant BSMOXFORD no. 228169. JK acknowledges support from the US DOE under contract no. DE-AC02-76SF00515.

\bibliographystyle{utphys}
\bibliography{LargeDBib}

\end{document}